\documentclass{article}%
\usepackage{amsmath}
\usepackage{color}
\usepackage{geometry}%
\usepackage{amsfonts}%
\usepackage{amssymb}%
\usepackage{graphicx}
\providecommand{\U}[1]{\protect\rule{.1in}{.1in}}
\newtheorem{theorem}{Theorem}
\newtheorem{acknowledgement}[theorem]{Acknowledgement}

\begin{document}

\title{Exact deflection of a Neutral-Tachyon in the Kerr's Gravitational field.}
\author{G. V. Kraniotis\thanks{Email:
{\color{blue}gkraniot@cc.uoi.gr}%
. }\\University of Ioannina, Department of Physics, GR 451 10 Ioannina, Greece}
\maketitle

\begin{abstract}
We solve in closed analytic form space-like geodesic equations in the Kerr
gravitational field. Such geodesic equations describe the motion of neutral
tachyons ( faster than light particles) in the Kerr spacetime. More
specifically we derive the closed form solution for the deflection angle of a
neutral tachyon on an equatorial orbit in Kerr spacetime. The solution is
expressed elegantly in terms of Lauricella's hypergeometric function $F_{D}%
.$We applied our results to three cases: first, for the calculation of the
deflection angle of a neutral tachyon on an equatorial trajectory in the
gravitational field of a Kerr black hole. Subsequently, we applied our exact
solutions to compute the deflection angle of equatorial spacelike geodesics in
the gravitational fields of Sun and Earth assuming the Kerr spacetime geometry.

\end{abstract}

\section{Introduction}

Recent results from the OPERA experiment, taken at face value, seem to
indicate the existence of superluminal neutrino species, whose speed
$\ \upsilon$ as it has been determined by OPERA collaboration with respect to
the speed of light, $c$ , is \cite{OPERA}:%
\begin{equation}
\frac{\upsilon-c}{c}=(2.48\pm0.28(\mathrm{stat)}\pm0.30(\mathrm{sys)}%
)\times10^{-5}\label{experiment}%
\end{equation}
for a mean energy of the neutrino beam of 17$%
\operatorname{GeV}%
\cite{OPERA}.$

This result, if it will stand to further scrutiny and will be confirmed by
independent measurements \ it will certainly constitute a fundamental
discovery about Nature.

For us it means, that we are endowed with sufficient motivation to explore
further properties and aspects of the general hypothesis of the existence of
superluminal (tachyon) particles in fundamental physics. In particular, we
intend in this Letter to explore the interaction of a neutral tachyon field
with the strong gravitational field of Kerr spacetime. We shall obtain the
exact analytic solution for the deflection angle of a neutral tachyon in an
equatorial unbound orbit in Kerr spacetime.

The plausible existence of a superluminal particle in the framework of special
relativity (SR)\ has been discussed by various authors \cite{BILANIUK} (see
also \cite{RECAMI}). The energy of a tachyon particle is given by
\cite{Ohanian}
\begin{equation}
E=\frac{|m|c^{2}}{\sqrt{\frac{\upsilon^{2}}{c^{2}}-1}},\text{ \textrm{for }%
}\upsilon>c
\end{equation}
where $|m|$ denotes the magnitude $\footnote{Sometimes refer to as the
\textit{metamass }\cite{BILANIUK}.}$ of the tachyon's imaginary rest mass:
$m=i|m|.$ The energy momentum \ vector is now%
\begin{equation}
p^{\mu}=\left(  \frac{|m|c}{\sqrt{\frac{\upsilon^{2}}{c^{2}}-1}},\frac
{|m|}{\sqrt{\frac{\upsilon^{2}}{c^{2}}-1}}\frac{\mathrm{d}x}{\mathrm{d}%
t},\frac{|m|}{\sqrt{\frac{\upsilon^{2}}{c^{2}}-1}}\frac{\mathrm{d}%
y}{\mathrm{d}t},\frac{|m|}{\sqrt{\frac{\upsilon^{2}}{c^{2}}-1}}\frac
{\mathrm{d}z}{\mathrm{d}t}\right)  ,\text{ \textrm{for }}\upsilon>c
\end{equation}
and the dispersion relation is valid%
\begin{equation}
p^{\mu}p_{\mu}=(E/c)^{2}-p_{x}^{2}-p_{y}^{2}-p_{z}^{2}=-|m|c^{2}.
\end{equation}

\bigskip

Assuming the gravitational mass of the tachyon is exactly equal to the
magnitude of its imaginary rest mass the tachyon moves along spacelike
geodesics in a gravitational field. Thus since spacelike geodesics are part of
the theory of General Relativity (GTR), tachyons can be considered as implicit
ingredients of the theory if the above assumption is valid. We mention at this
point earlier works on this matter. The author of \cite{Sum} has performed a
weak field calculation of the deflection angle of a neutral tachyon in the
gravitational field of the Sun assuming a Schwarzschild spacetime geometry.
Tachyons in uniform relativistic cosmology have been considered in
\cite{Schwarz} and an initial study of the conditions for tachyons to be
captured by a Kerr black hole has been performed in \cite{Narlikar}. However,
an exact analytic calculation of the deflection angle for an equatorial orbit
in the Kerr gravitational field is still lacking. Our Letter fills this
important gap in the field. We start our discussion with the derivation of
spacelike geodesics in the Kerr spacetime with a cosmological constant present
and subsequently we obtain the closed form analytic solution for the
deflection angle of an unbound tachyonic equatorial orbit in the Kerr
gravitational field . The exact solution is expressed in terms of generalized
hypergeometric functions of Appell-Lauricella. We apply our closed form
solutions to three cases. We determine the deflection angle that an equatorial
spacelike geodesic undergoes in the gravitational fields of the Sun and Earth
assuming a Kerr geometry for modelling these fields. We also apply our
solutions to compute the exact deflection angle that a neutral tachyon on an
equatorial trajectory undergoes in the gravitational field exerted by a
rotating (Kerr) black hole. We leave the last section for our conclusions.

\section{Spacelike geodesics in Kerr-de Sitter spacetime.}

\bigskip Taking into account the contribution from the cosmological constant
$\Lambda$, the generalization of the Kerr solution \cite{Kerr} is described by
the Kerr-de Sitter metric element which in Boyer-Lindquist (BL) coordinates is
given by \cite{Stuchlik}-\cite{Demianski}:

\bigskip%

\begin{align}
\mathrm{d}s^{2}  & =\frac{\Delta_{r}}{\Xi^{2}\rho^{2}}(c\mathrm{d}t-a\sin
^{2}\theta\mathrm{d}\phi)^{2}-\frac{\rho^{2}}{\Delta_{r}}\mathrm{d}r^{2}%
-\frac{\rho^{2}}{\Delta_{\theta}}\mathrm{d}\theta^{2}\nonumber\\
& -\frac{\Delta_{\theta}\sin^{2}\theta}{\Xi^{2}\rho^{2}}(ac\mathrm{d}%
t-(r^{2}+a^{2})\mathrm{d}\phi)^{2}%
\end{align}%
\begin{equation}
\Delta_{\theta}:=1+\frac{a^{2}\Lambda}{3}\cos^{2}\theta,\text{ \ \ }%
\Xi:=1+\frac{a^{2}\Lambda}{3}%
\end{equation}

\begin{equation}
\Delta_{r}:=\left(  1-\frac{\Lambda}{3}r^{2}\right)  \left(  r^{2}%
+a^{2}\right)  -2\frac{GM}{c^{2}}r,\text{ \ \ }\rho^{2}=r^{2}+a^{2}\cos
^{2}\theta\label{AktinDr}%
\end{equation}

We denote by $a$ the rotation (Kerr) parameter and $M$ denotes the mass of the
spinning black hole. Choosing a real affine parameter $\lambda$ for the
spacelike geodesic by $\mathrm{d}\lambda^{2}=-\mathrm{d}s^{2},$ we have the
geodesic equation in the usual notation as%
\begin{equation}
\frac{\mathrm{d}^{2}x^{i}}{\mathrm{d}\lambda^{2}}+\Gamma_{jk}^{i}%
\frac{\mathrm{d}x^{j}}{\mathrm{d}\lambda}\frac{\mathrm{d}x^{k}}{\mathrm{d}%
\lambda}=0,\label{geodesia}%
\end{equation}
where $x^{i}$ denote the BL coordinates and $\Gamma_{jk}^{i}$ the Christoffel
symbols of the second kind.

The geodesic equations in Kerr spacetime in the presence of the cosmological
constant $\Lambda$ were derived in \cite{Kraniotis1} by solving the
Hamilton-Jacobi differential equations by separation of variables. The tachyon
motion, as we mentioned earlier, is described by the spacelike geodesics
(which are the first integrals of (\ref{geodesia})) which take the
form\footnote{The tachyon metamass $|m|,$ which is one of the first integrals
of (\ref{geodesia}), was set equal to unity without loss of generality.}:%
\begin{align}
\int\frac{\mathrm{d}r}{\sqrt{R}}  & =\int\frac{\mathrm{d}\theta}{\sqrt{\Theta
}},\nonumber\\
\rho^{2}\frac{\mathrm{d}\phi}{\mathrm{d}\lambda}  & =-\frac{\Xi^{2}}%
{\Delta_{\theta}\sin^{2}\theta}(aE\sin^{2}\theta-L)+\frac{a\Xi^{2}}{\Delta
_{r}}[(r^{2}+a^{2})E-aL]\nonumber\\
c\rho^{2}\frac{\mathrm{d}t}{\mathrm{d}\lambda}  & =\frac{\Xi^{2}(r^{2}%
+a^{2})[(r^{2}+a^{2})E-aL]}{\Delta_{r}}-\frac{a\Xi^{2}(aE\sin^{2}\theta
-L)}{\Delta_{\theta}}\label{spaceKdS}\\
\rho^{2}\frac{\mathrm{d}r}{\mathrm{d}\lambda}  & =\pm\sqrt{R}\nonumber\\
\rho^{2}\frac{\mathrm{d}\theta}{\mathrm{d}\lambda}  & =\pm\sqrt{\Theta
}\nonumber
\end{align}

where%
\begin{align}
R  & :=\Xi^{2}[(r^{2}+a^{2})E-aL]^{2}-\Delta_{r}(-r^{2}+Q+\Xi^{2}%
(L-aE)^{2})\nonumber\\
\Theta & :=[Q+(L-aE)^{2}\Xi^{2}+a^{2}\cos^{2}\theta]\Delta_{\theta}-\Xi
^{2}\frac{(aE\sin^{2}\theta-L)^{2}}{\sin^{2}\theta}\label{SLKDS}%
\end{align}
The constants of motion $E,$ $L$ are associated with the
isometries\footnote{i.e. they are related to the energy and angular momentum
per unit metamass of the tachyon at infinity.} of the Kerr metric while $Q$
denotes Carter's constant, the fourth constant \ of integration. The spacelike
Kerr geodesics are obtained by setting $\Lambda=0$ in Eqs.(\ref{spaceKdS}%
)-(\ref{SLKDS}).

\section{Exact solution of equatorial spacelike geodesics in Kerr spacetime}

We now proceed to determine the exact solution for the deflection of a neutral
tachyon in an equatorial spacelike orbit in Kerr spacetime assuming
$\Lambda=0$. The more general case with the cosmological constant present will
be a subject of a separate publication \cite{KraniotisTach}. We have
$r^{2}(\dot{r})=\sqrt{R}.$This can be rewritten as \footnote{For equatorial
geodesics $\theta=\pi/2,$ $Q=0.$}
\begin{equation}
\dot{r}^{2}=E^{2}+\frac{a^{2}E^{2}}{r^{2}}-\frac{L^{2}}{r^{2}}+\frac
{2GM}{c^{2}r^{3}}(L-aE)^{2}+\frac{\Delta}{r^{2}}.
\end{equation}
where $\Delta$ is obtained by setting $\Lambda=0$ in equation (\ref{AktinDr})
for $\Delta_{r}$. By defining a new variable $u:=1/r$ we obtain the following
expression:%
\begin{equation}
u^{-4}\dot{u}^{2}=E^{2}+a^{2}E^{2}u^{2}-L^{2}u^{2}+\frac{2GM}{c^{2}}%
u^{3}(L-aE)^{2}+(1+a^{2}u^{2}-\frac{2GM}{c^{2}}u)\equiv B_{tac}(u).
\end{equation}
Similarly the geodesic for the azimuth coordinate is given%
\begin{equation}
\dot{\phi}^{2}=u^{4}\frac{A^{2}(u)}{D^{2}(u)}%
\end{equation}
where%
\begin{equation}
A(u):=L+\alpha_{S}u(aE-L),\qquad D(u):=1+a^{2}u^{2}-\alpha_{S}u,\qquad
\alpha_{S}:=\frac{2GM}{c^{2}}.
\end{equation}
Thus we derive the orbital equation%
\begin{equation}
\frac{\mathrm{d}\phi}{\mathrm{d}u}=\frac{A(u)}{D(u)}\frac{1}{\sqrt{B_{tac}%
(u)}}.\label{Orbitalequation}%
\end{equation}
We now use the technique of partial fractions from integral calculus in order
to calculate the deflection of the neutral tachyon's equatorial orbit in the
Kerr gravitational field from equation (\ref{Orbitalequation}). We write:%
\begin{equation}
\frac{A(u)}{D(u)}=\frac{A_{+}}{u_{+}-u}+\frac{A_{-}}{u_{-}-u}%
\end{equation}
where $u_{+}=\frac{r_{+}}{a^{2}},u_{-}=\frac{r_{-}}{a^{2}}$ and%
\begin{equation}
r_{\pm}=\frac{GM}{c^{2}}\pm\sqrt{\left(  \frac{GM}{c^{2}}\right)  ^{2}-a^{2}}%
\end{equation}
denote the radii of the event and Cauchy horizons respectively for the case of
a Kerr black hole. Also the quantities $A_{+},A_{-}$ are given by%
\begin{equation}
A_{+}=\frac{\frac{L}{a^{2}}+\frac{\alpha_{S}}{a^{2}}(aE-L)u_{+}}{u_{-}-u_{+}%
}\qquad A_{-}=\frac{\frac{-L}{a^{2}}-\frac{\alpha_{S}}{a^{2}}(aE-L)u_{-}%
}{u_{-}-u_{+}}%
\end{equation}
In order to calculate the angle of deflection for the tachyon it is necessary
to calculate the integral: $\Delta\phi_{Tachyon}^{GTR}=2\int_{0}%
^{u_{2}^{\prime}}\mathrm{d}\phi.$Using the formalism developed in references
\cite{Kraniotis1}-\cite{KraniotisGravLens} , for computing hyperelliptic
integrals in closed analytic form, in terms of Lauricella's hypergeometric
function $F_{D}$ , we compute:
\begin{align}
\Delta\phi_{Tachyon}^{GTR}  & =\frac{2}{\sqrt{u_{1}^{\prime}-u_{3}^{\prime}}%
}\frac{1}{\sqrt{\frac{\alpha_{S}(L-aE)^{2}}{(\frac{GM}{c^{2}})^{3}}}}%
\Biggl\{%
\frac{A_{+}}{\frac{GMr_{+}}{c^{2}a^{2}}-u_{3}^{\prime}}%
\Biggl(%
F_{1}\left(  \frac{1}{2},\mathbf{\beta}_{A},1,\mathbf{z}_{A}^{r_{+}}\right)
\pi\nonumber\\
& -2\sqrt{\frac{-u_{3}^{\prime}}{u_{2}^{\prime}-u_{3}^{\prime}}}F_{D}\left(
\frac{1}{2},\mathbf{\beta}_{3}^{4},\frac{3}{2},\mathbf{z}_{D}^{r_{+}}\right)
\Biggr)%
\nonumber\\
& +\frac{A_{-}}{\frac{GMr_{-}}{c^{2}a^{2}}-u_{3}^{\prime}}%
\Biggl(%
F_{1}\left(  \frac{1}{2},\mathbf{\beta}_{A},1,\mathbf{z}_{A}^{r_{-}}\right)
\pi\nonumber\\
& -2\sqrt{\frac{-u_{3}^{\prime}}{u_{2}^{\prime}-u_{3}^{\prime}}}F_{D}\left(
\frac{1}{2},\mathbf{\beta}_{3}^{4},\frac{3}{2},\mathbf{z}_{D}^{r_{-}}\right)
\Biggr)%
\Biggr\}%
\label{DeflectionTachyon}%
\end{align}
where
\begin{equation}
\mathbf{\beta}_{A}=\left(  1,\frac{1}{2}\right)  ,\quad\mathbf{\beta}_{3}%
^{4}=\left(  1,\frac{1}{2},\frac{1}{2}\right)
\end{equation}
and
\begin{equation}
\mathbf{z}_{A}^{r_{\pm}}=\left(  \frac{u_{2}^{\prime}-u_{3}^{\prime}}%
{\frac{GMr_{\pm}}{c^{2}a^{2}}-u_{3}^{\prime}},\frac{u_{2}^{\prime}%
-u_{3}^{\prime}}{u_{1}^{\prime}-u_{3}^{\prime}}\right)  ,\quad\mathbf{z}%
_{D}^{r_{\pm}}=\left(  \frac{-u_{3}^{\prime}}{\frac{GMr_{\pm}}{c^{2}a^{2}%
}-u_{3}^{\prime}},\frac{-u_{3}^{\prime}}{u_{1}^{\prime}-u_{3}^{\prime}}%
,\frac{-u_{3}^{\prime}}{u_{2}^{\prime}-u_{3}^{\prime}}\right)
\end{equation}
We have also defined: $u^{\prime}=u\frac{GM}{c^{2}}$ . The roots of the cubic
in this case are real and organized in the ascending order
\begin{equation}
u_{1}^{\prime}>u_{2}^{\prime}>0>u_{3}^{\prime}%
\end{equation}

The angle of deflection $\delta$ of a neutral tachyon equatorial trajectory
from the gravitational field of a rotating black hole or a rotating central
mass is defined to be the deviation of $\Delta\phi_{Tachyon}^{GTR}$ from the
transcendental number $\pi$%
\begin{equation}
\delta=\Delta\phi_{Tachyon}^{GTR}-\pi.\label{DeflePI}%
\end{equation}

We shall apply our closed form solution, Eq.(\ref{DeflectionTachyon}) for the
deflection angle of an equatorial neutral tachyon's orbit in the gravitational
field of Kerr spacetime \ in two cases. First, we shall compute the deflection
angle $\delta,$ for a tachyon in the gravitational field of the Sun (assuming
the Kerr spacetime geometry). Second, the deflection angle of a neutral
tachyon in an equatorial orbit around a Kerr black hole for various values of
the involved physical parameters. The physical parameters are the velocity of
the tachyon particle $\upsilon$ (at large distances from the central mass)$,$
the energy per unit mass (in units of $c^{2})$ $E=\frac{1}{\sqrt{\upsilon
^{2}/c^{2}-1}}$ , the parameter $L$ and the spin $a$ (Kerr parameter) of the
black hole$.$

\subsection{Deflection of neutral tachyon in an equatorial orbit in the Kerr
(Sun's) gravitational field.}

Assuming that the gravitational field of the spinning Sun is described by the
Kerr spacetime geometry we \ compute the deflection angle of a tachyon in an
unbound equatorial orbit using our closed form solution Eq.(\ref{DeflePI}). We
repeat the calculation for three values of the velocity of the tachyon
particle. Namely: 1) $\upsilon=(1+2.48\times10^{-5})c$ $($same as OPERA's
experimental value) 2) $\upsilon=\sqrt{2}c$ 3) $\upsilon=10^{6}c.$ For the
\ parameter $E$ we use the formula $E=\frac{1}{\sqrt{\upsilon^{2}/c^{2}-1}}.$
Taking the point of closest approach $r_{0}=R_{\odot}=6.9551\times10^{8}%
\operatorname{m}%
$ the parameter $L$ is chosen to have the value: $L=471013.2781620\upsilon
E\frac{G_{N}M_{\odot}}{c^{2}}.$Our results are summarized in Table
\ref{TachyonSun}. We note that the deflection angle ranges from half the value
of a photon's deflection from the Sun's gravitational field (case 3 high
velocitites, low energies for the tachyon) to a value equal to the deflection
of light for the case 1) i.e. tachyon velocity as the one determined by OPERA
experiment (\ref{experiment}).%

\begin{table}[tbp] \centering
\begin{tabular}
[c]{|l|l|l|}\hline
Case1: $\upsilon=(1+2.48\times10^{-5})c$ & Case 2: $\upsilon=\sqrt{2}c$ & Case
3: $\upsilon=10^{6}c$\\\hline
$\delta=1.75164\mathrm{\operatorname{arcsec}}$ & $\delta
=1.31376\mathrm{\operatorname{arcsec}}$ & $\delta
=0.875837\mathrm{\operatorname{arcsec}}$\\
&  & \\
&  & \\\hline
\end{tabular}
\caption{Deflection of a neutral tachyon's equatorial trajectory in the
gravitational field of the Sum assuming Kerr spacetime geometry. The value
of the Kerr parameter was chosen to be $a=0.2158\frac{GM_{\odot}}{c^2}$.}\label{TachyonSun}%
\end{table}%

Our findings are consistent with the results of \cite{Sum}, in which, a
perturbative calculation of the deflection angle for a tachyon in a
Schwarzschild field of the Sun was performed.

\subsection{Deflection of a neutral tachyon in an equatorial orbit in the Kerr
black hole spacetime}

Again for different choices of values for the velocity of a tachyon we compute
the deflection angle for different values for the spin of the black hole and
the remaining parameters. For the three different values of the tachyon
particle we choose first $L=100\upsilon E\frac{GM_{\mathrm{BH}}}{c^{2}}.$ We
present our results in table \ref{BlackHoleKerr}.%

\begin{table}[tbp] \centering
\begin{tabular}
[c]{|l|l|l|l|}\hline
$a$ & $\upsilon=(1+2.48\times10^{-5})c$ & $\upsilon=\sqrt{2}c$ &
$\upsilon=10^{6}c$\\\hline
$0.52$ & $\delta=0.04099735=8456.31\operatorname{arcsec}$ & $\delta
=0.0305724=6306.01\operatorname{arcsec}$ & $\delta
=0.0202396=4174.71\operatorname{arcsec}$\\
$0.99616$ & $\delta=0.04079425=8414.42\operatorname{arcsec}$ & $\delta
=0.0304312=6276.89\operatorname{arcsec}$ & $\delta
=0.02024108=4175.02\operatorname{arcsec}$\\
&  &  & \\\hline
\end{tabular}
\caption{Deflection angle of a neutral tachyon in an equatorial orbit in the
gravitational field of a Kerr black hole. We present our results for three
different values of the velocity of the tachyon particle and two choices for
the spin of the black hole. Also $L=100\upsilon E \frac{GM_{\rm BH}}{c^2}$
and $E=\frac{1}{\sqrt{\upsilon^2 / c^2-1}}$.}\label{BlackHoleKerr}%
\end{table}%

We repeat the analysis for $L=40\upsilon E\frac{GM_{\mathrm{BH}}}{c^{2}}.$ Our
results are dislayed in Table \ref{BlackHoleKerrLowL}.%

\begin{table}[tbp] \centering
\begin{tabular}
[c]{|l|l|l|l|}\hline
$a$ & $\upsilon=(1+2.48\times10^{-5})c$ & $\upsilon=\sqrt{2}c$ &
$\upsilon=10^{6}c$\\\hline
$0.52$ & $\delta=0.10652474=21972.3\operatorname{arcsec}$ & $\delta
=0.0787036=16233.8\operatorname{arcsec}$ & $\delta
=0.05153626=10630.1\operatorname{arcsec}$\\
$0.99616$ & $\delta=0.10512217=21683\operatorname{arcsec}$ & $\delta
=0.0777545=16038\operatorname{arcsec}$ & $\delta
=0.05156155=10635.3\operatorname{arcsec}$\\
&  &  & \\\hline
\end{tabular}
\caption{Deflection angle of a neutral tachyon in an equatorial orbit in the
gravitational field of a Kerr black hole. We present our results for three
different values of the velocity of the tachyon particle and two choices for
the spin of the black hole. Also $L=40\upsilon E \frac{GM_{\rm BH}}{c^2}$
and $E=\frac{1}{\sqrt{\upsilon^2 / c^2-1}}$.}\label{BlackHoleKerrLowL}%
\end{table}%

\subsection{Deflection of a neutral tachyon in an equatorial orbit by the
gravitational field of Earth}

Assuming the Earth's gravitational field is described by a Kerr spacetime, we
shall compute the deflection angle of a neutral tachyon in an equatorial orbit
in Earth's gravitation. There is a further novelty in this calculation. \ The
Kerr parameter that corresponds to the angular momentum of Earth is equal to
$a_{\oplus}=329.432%
\operatorname{cm}%
=371.398(2GM_{\oplus}/c^{2})\cite{KraniotisMPI}.$ Thus the two roots of the
polynomial $D(u)$ are complex-conjugate. Therefore, when we calculate exactly
the hyperelliptic integral, two of the variables of Lauricella's function
$F_{D}$ are complex-conjugates. This is fine as long as their modules are less
than $1,$which indeed it is the case in our computations.

Thus%
\begin{align}
\int\mathrm{d}\phi & =2\int_{0}^{u_{2}^{\prime}}\frac{A(u)}{D(u)}\frac
{1}{\sqrt{B_{tac}(u)}}\mathrm{d}u\nonumber\\
& =2\int_{0}^{u_{2}^{\prime}}\frac{\mathrm{d}u^{\prime}L}{a^{2}\left(
\frac{GM_{\oplus}r_{+}}{c^{2}a^{2}}-u^{\prime}\right)  \left(  \frac
{GM_{\oplus}r_{-}}{c^{2}a^{2}}-u^{\prime}\right)  }\frac{1}{\sqrt{\frac
{\alpha_{S}(L-aE)^{2}}{(GM_{\oplus}/c^{2})^{3}}}}\frac{1}{\sqrt{(u^{\prime
}-u_{3}^{\prime})(u_{1}^{\prime}-u^{\prime})(u_{2}^{\prime}-u^{\prime})}%
}+\nonumber\\
& 2\int_{0}^{u_{2}^{\prime}}\frac{\alpha_{S}(aE-L)\mathrm{d}u^{\prime
}u^{\prime}}{a^{2}\left(  \frac{GM_{\oplus}r_{+}}{c^{2}a^{2}}-u^{\prime
}\right)  \left(  \frac{GM_{\oplus}r_{-}}{c^{2}a^{2}}-u^{\prime}\right)
\sqrt{\frac{\alpha_{S}(L-aE)^{2}}{(GM_{\oplus}/c^{2})^{3}}}\sqrt{(u^{\prime
}-u_{3}^{\prime})(u_{1}^{\prime}-u^{\prime})(u_{2}^{\prime}-u^{\prime})}}%
\end{align}
Now applying the transformation
\begin{equation}
\fbox{$\displaystyle u^{\prime}=u_2^{\prime}(1-t),
\nonumber$}
\end{equation}
we have:%
\begin{align}
\frac{GM_{\oplus}r_{\pm}}{c^{2}a^{2}}-u^{\prime}  & =\left(  \frac{GM_{\oplus
}r_{\pm}}{c^{2}a^{2}}-u_{2}^{\prime}\right)  \left[  1+\frac{tu_{2}^{\prime}%
}{\frac{GM_{\oplus}r_{\pm}}{c^{2}a^{2}}-u_{2}^{\prime}}\right]  ,\text{
\ }u_{2}^{\prime}-u^{\prime}=u_{2}^{\prime}t,\\
u_{1}^{\prime}-u^{\prime}  & =(u_{1}^{\prime}-u_{2}^{\prime})\left[
1+\frac{u_{2}^{\prime}t}{u_{1}^{\prime}-u_{2}^{\prime}}\right]  ,\text{
\ }u^{\prime}-u_{3}^{\prime}=(u_{2}^{\prime}-u_{3}^{\prime})\left[
1-\frac{u_{2}^{\prime}t}{u_{2}^{\prime}-u_{3}^{\prime}}\right]
\end{align}
thus we tranform our integral onto the integral representation of Lauricella's
function $F_{D}$ of four-variables (see also Appendix):
\begin{align}
\int\mathrm{d}\phi & =\Delta\phi_{NT\oplus}^{GTR}\nonumber\\
& =\frac{2}{a^{2}\left(  \frac{GM_{\oplus}r_{+}}{c^{2}a^{2}}-u_{2}^{\prime
}\right)  \left(  \frac{GM_{\oplus}r_{-}}{c^{2}a^{2}}-u_{2}^{\prime}\right)
\sqrt{u_{2}^{\prime}(u_{2}^{\prime}-u_{3}^{\prime})(u_{1}^{\prime}%
-u_{2}^{\prime})}\sqrt{\frac{\alpha_{S}(L-aE)^{2}}{(GM_{\oplus}/c^{2})^{3}}}%
}\times\nonumber\\
&
\Biggl[%
\alpha_{S}(aE-L)u_{2}^{\prime2}\frac{\Gamma(1/2)\Gamma(2)}{\Gamma(5/2)}%
F_{D}\left(  \frac{1}{2},\mathbf{\beta}_{4}^{1},\frac{5}{2},\mathbf{z}%
_{r}^{\oplus}\right)  +\nonumber\\
& Lu_{2}^{\prime}\frac{\Gamma(1/2)\Gamma(1)}{\Gamma(3/2)}F_{D}\left(  \frac
{1}{2},\mathbf{\beta}_{4}^{1},\frac{3}{2},\mathbf{z}_{r}^{\oplus}\right)
\Biggr]%
\label{MotherEarth}%
\end{align}
where
\begin{equation}
\mathbf{\beta}_{4}^{1}=\left(  1,1,\frac{1}{2},\frac{1}{2}\right)  ,\text{
\ \ }\mathbf{z}_{r}^{\oplus}=\left(  \frac{-u_{2}^{\prime}}{\frac{GM_{\oplus
}r_{+}}{c^{2}a^{2}}-u_{2}^{\prime}},\frac{-u_{2}^{\prime}}{\frac{GM_{\oplus
}r_{-}}{c^{2}a^{2}}-u_{2}^{\prime}},\frac{u_{2}^{\prime}}{u_{2}^{\prime}%
-u_{3}^{\prime}},\frac{-u_{2}^{\prime}}{u_{1}^{\prime}-u_{2}^{\prime}}\right)
\end{equation}
We apply our analytic solution eq.(\ref{MotherEarth}) for computing the
deflection angle of a neutral tachyon in the gravitational field of Earth. We
choose as value for the velocity of the tachyon the central value of OPERA
experiment $\upsilon=(1+2.48\times10^{-5})c.$ Taking the point of closest
approach as the radius (equatorial) of Earth $r_{0}=R_{\oplus}=6.378137\times
10^{6}%
\operatorname{m}%
$ we have that the parameter $L=\upsilon E\times1.438127796068894\times
10^{9}.$ For the Kerr parameter of Earth we choose the above mentioned value.
Then we determine $\delta=\Delta\phi_{NT\oplus}^{GTR}-\pi=2.78\times10^{-9}%
\operatorname{rad}%
\sim0.000573\operatorname{arcsec}.$

\section{Conclusions}

In this paper we investigated tachyon orbits (spacelike geodesics) in Kerr
spacetime. More specifically, we derived the closed form solution for the
deflection angle of a neutral tachyon in an equatorial orbit in the
gravitational field of Kerr spacetime. The solution was expressed elegantly in
terms of Lauricella's multivariable hypergeometric function $F_{D}.$We applied
our exact solutions in three cases: 1) we calculated the deflection of a
neutral tachyon by the gravitational field of a rotating Kerr black hole, for
different values of the velocity of the tachyon and the spin of the black
hole. We note the strong dependence of the deflection angle on the spin of the
spinning black hole for low tachyon velocities, especially for lower \ values
for the parameter $L$. Large magnitudes for the deflection angle were produced
see table \ref{BlackHoleKerrLowL}. 2) we calculated the deflection of
equatorial neutral tachyon orbits by the gravitational field of our Sun
assuming a curved Kerr spacetime geometry. For low tachyon velocities (such as
the ones reported by the OPERA collaboration) the deflection angle was
calculated to have a value $\sim1.75\operatorname{arcsec}$ a value equals to
the amount of deflection that light experiences by the gravitation field of
our Solar system star. For high tachyon velocities $\upsilon\gg c$ the
calculated deflection angle decreases to half the value of $\delta$ at low
superluminal velocities. \ 3) we calculated the deflection angle of an
equatorial tachyon trajectory in the gravitational field of Earth assuming a
Kerr geometry. There is a further novelty in the calculation. The solution for
$\delta$ is expressed in terms of Lauricella's hypergeometric function $F_{D}$
of four variables two of which are complex-conjugates. The neutral tachyon
undergoes a small deflection of $2.78\times10^{-9}$ radians$\sim
0.000573\operatorname{arcsec}$. Thus if tachyons do exist and move on
spacelike geodesics they undergo a deflection by the gravitational field of
the rotating central mass. The deflection exhibits a strong dependence on the
superluminal velocity and the spin of the rotating mass. This gravitational
effect is in principle measurable. It will be interesting to generalize our
results to the case of finding the exact solutions for generic non-equatorial
spacelike orbits in the presence of the cosmological constant $\Lambda$
(spacelike non-equatorial Kerr-de Sitter orbits). \ However, such an
investigation is beyond the scope of the current paper and it will be the
subject of a separate publication \cite{KraniotisTach}.

\begin{acknowledgement}
The author acknowledges useful discussions with Giota Grigoriadou.
\end{acknowledgement}

\section{Appendix}

We introduce the Lauricella function $F_{D}$ and its integral representation%

\begin{equation}
\fbox{$\displaystyle F_D(\alpha,{\bf\beta},\gamma,{\bf z})=
\sum_{n_1,n_2,\dots,n_m=0}^{\infty}\frac{(\alpha)_{n_1+\cdots n_m}%
(\beta_1)_{n_1} \cdots(\beta_m)_{n_m}}
{(\gamma)_{n_1+\cdots+n_m}(1)_{n_1}\cdots(1)_{n_m}} z_1^{n_1}\cdots
z_m^{n_m}$}
\label{GLauri}
\end{equation}%
where
\begin{align}
\mathbf{z}  & =(z_{1},\ldots,z_{m}),\nonumber\\
\mathbf{\beta}  & =(\beta_{1},\ldots,\beta_{m}).\label{GLFD}%
\end{align}

The Pochhammer symbol
\fbox{$\displaystyle(\alpha)_m=(\alpha,m)$}
is defined by%
\begin{equation}
(\alpha)_{m}=\frac{\Gamma(\alpha+m)}{\Gamma(\alpha)}=\left\{
\begin{array}
[c]{ccc}%
1, &
{\rm if}%
& m=0\\
\alpha(\alpha+1)\cdots(\alpha+m-1) & \text{%
{\rm if}%
} & m=1,2,3
\end{array}
\right.
\end{equation}
With the notations $\mathbf{z}^{\mathbf{n}}:=z_{1}^{n}\ldots z_{m}^{n},$
$(\mathbf{\beta)}_{\mathbf{n}}:=(\beta_{1})_{n_{1}}\ldots(\beta_{m})_{n_{m}},$
$\mathbf{n!=}n_{1}!\ldots n_{m}!,|\mathbf{n}|:=n_{1}+\ldots n_{m}$ for
m-tuples of numbers in (\ref{GLFD}) and of non-negative integers
$\mathbf{n=}(n_{1},\ldots n_{m})$ the Lauricella series $F_{D}$ in compact
form is%
\begin{equation}
F_{D}(\alpha,\mathbf{\beta,}\gamma,\mathbf{z):=}\sum_{\mathbf{n}}\frac
{(\alpha)_{|\mathbf{n}|}(\mathbf{\beta})_{\mathbf{n}}}{(\gamma)_{|\mathbf{n}%
|}\mathbf{n!}}\mathbf{z}^{\mathbf{n}}.
\end{equation}
The series admits the following integral representation:%

\begin{equation}
\fbox{$\displaystyle F_D(\alpha,{\bf\beta},\gamma,{\bf z})=
\frac{\Gamma(\gamma)}{\Gamma(\alpha)\Gamma(\gamma-\alpha)}
\int_0^1 t^{\alpha-1}(1-t)^{\gamma-\alpha-1}(1-z_1 t)^{-\beta_1}%
\cdots(1-z_m t)^{-\beta_m} {\rm d}t $}
\label{OloklAnapa}
\end{equation}%

which is valid for
\fbox{$\displaystyle{\rm Re}(\alpha)>0,\;{\rm Re}(\gamma-\alpha)>0$}%
. It
{\em converges\;absolutely}
inside the m-dimensional cuboid:%
\begin{equation}
|z_{j}|<1,(j=1,\ldots,m).
\end{equation}
For $m=2,$ $F_{D}$ becomes Appell's $F_{1}$ two variable hypergeometric
function $F_{1}(\alpha,\beta,\beta^{\prime},\gamma,x,y)$ with integral
representation%
\begin{equation}
\int_{0}^{1}u^{\alpha-1}(1-u)^{\gamma-\alpha-1}(1-ux)^{-\beta}(1-yu)^{-\beta
^{\prime}\mathrm{\ }}\mathrm{d}u=\frac{\Gamma(\alpha)\Gamma(\gamma-\alpha
)}{\Gamma(\gamma)}F_{1}(\alpha,\beta,\beta^{\prime},\gamma,x,y).
\end{equation}

\end{document}